\documentclass[
]{ws-ijtaf}

\usepackage{dsfont}
\usepackage{natbib}
\usepackage[caption=false]{subfig}
\usepackage{graphicx}
\usepackage{color}

\begin{document}
\title{Quantile correlations: Uncovering temporal dependencies in financial time series}
\author{THILO A. SCHMITT}
\address{
Fakult\"at f\"ur Physik, Universit\"at Duisburg-Essen\\
47048 Duisburg, Germany\\
\email{thilo.schmitt@uni-due.de}}

\author{RUDI SCH\"AFER}
\address{
Fakult\"at f\"ur Physik, Universit\"at Duisburg-Essen\\
47048 Duisburg, Germany\\
}
\author{HOLGER DETTE}
\address{
Fakult\"at f\"ur Mathematik, Ruhr-Universit\"at Bochum\\
44780 Bochum, Germany\\
}
\author{THOMAS GUHR}
\address{
Fakult\"at f\"ur Physik, Universit\"at Duisburg-Essen\\
47048 Duisburg, Germany\\
}


\maketitle

\begin{abstract}We conduct an empirical study using the quantile-based correlation function to uncover the temporal dependencies in financial time series. The study uses intraday data for the S\&P 500 stocks from the New York Stock Exchange. After establishing an empirical overview we compare the quantile-based correlation function to stochastic processes from the GARCH family and find striking differences. This motivates us to propose the quantile-based correlation function as a powerful tool to assess the agreements between stochastic processes and empirical data.
\end{abstract}


\section{Introduction}

Financial time series exhibit various non-trivial properties. The distribution of returns deviates from a normal distribution and shows heavy tails~\citep{Oliver1926,Mills1927,Mandelbrot1963}. This behavior was first observed by~\cite{Mitchell1915}. In stochastic volatility models the normal distribution is combined with a distribution for the variances. The resulting return distribution shows non-normality, see \textit{e.g.}, \cite{Clark1973} and \cite{Yang2004}. However, simply drawing volatilities from a distribution does not account for the empirically observed volatility clustering, \textit{i.e.}, temporal inhomogeneity. To achieve this the description by stochastic processes is essential. In his groundbreaking work Engle made great strides towards this goal by introducing the ARCH process, see~\cite{Engle1982}. 

The autocorrelation of the return time series is zero~\citep{Pagan1996}. However, the autocorrelation of the time series for absolute or squared returns is non-zero and slowly decays to zero for larger lags~\citep{Ding1993,Cizeau1997,Liu1997}. This effect can be traced back to volatility clustering~\cite{Mandelbrot1963}. In high volatility phases the probability that a large (absolute) return is followed by another large return is higher than normal. The same holds true for phases of small volatility, where small returns are followed by small returns with higher probability. Closely related is the so called ``leverage effect''. It is empirically known that volatilities and returns show a negative correlation. This was first observed by~\cite{Black:1976fj} and attributed to the fact that stocks with falling prices are riskier and therefore have a higher volatility. The reduced market capitalization relative to the debt of the company makes it more leveraged, hence the name. This explanation is often challenged in the literature~\citep{Figlewski2000,Aydemir2006,Ait-Sahalia2013}. The studies of these effects typically focuses on the autocorrelation of return time series or cross correlations between returns and historical or implied volatilities.

A common tool to analyze temporal dependencies in time series is the $L^2$-periodogram, which is intrinsically connected to mean values and covariances 
and has several optimality properties for  the analysis of Gaussian series. On the other hand it is well known that this periodogram has difficulties to detect nonlinear dynamics 
such as changes in the conditional  shape (skewness, kurtosis) or  heavy tails, and several modifications  have been proposed to address these problems.
Laplace periodograms  have been investigated  as  an alternative  to the ordinary periodogram
  by~\cite{Li2008,Li2012}
  for analyzing heart rate  variability and sun spot data, in the frequency domain. These peridograms are based on  quantile regression methodology, see~\cite{Koenker2005} and extensions, which are independent with respect to monotone 
  transformation of the data  have been developed by~\cite{Dette2011a} and \cite{Hagemann2011}. These authors propose a different periodogram, which is defined as the discrete Fourier transform
  of quantile-based correlations, see~\cite{Kedem1980} and develop a corresponding statistical theory.
     
Here, we want to show that even the direct use of quantile-based correlations  provides  a very powerful tool to investigate nonlinear dynamics of  financial  series in the time domain. For this purpose
we conduct an empirical study on intraday data from stocks in the Standard \& Poor's 500-index (S\&P500) during the years 2007 and 2008. In addition, we show that quantile-based correlation is able to uncover subtle differences between stochastic processes and empirical data. As an example, we study the return time series from GARCH~\citep{Bollerslev1986}, EGARCH~\citep{Nelson1991} and GJR-GARCH~\citep{Glosten1993} processes.

\section{Quantile-based correlation}
\label{ch:qcf}

Given a time series $x=(x_1, x_2, \dots, x_T)$ of length $T$ we calculate the quantile-based correlation in the following way. Let $\alpha \in [0,1]$ and $\beta \in [0,1]$ be probability levels. Then, let $q_{\alpha}$ be the value at the $\alpha$-quantile for the time series $x$. We first map the time series $x$ to a filtered time series $\xi^{(\alpha)}$ according to a filter rule
\begin{equation}
\xi_t^{(\alpha)} =  \left\{ \begin{array}{l} 1 \quad , \quad x_t \leq q_{\alpha}\\	0 \quad , \quad {\rm otherwise}  \end{array} \right. 
	\quad .
	\label{eq:qcf}
\end{equation}
For example, if the time series is
\begin{equation}
	x=( 1, -5, 10, 0, -6, -2, -2, 2, 0, 2 )
\end{equation}
we have $q_{0.5}=0$ for the 0.5-quantile and the filtered time series is
\begin{equation}
	\xi^{(0.5)}=( 0, 1, 0, 1, 1, 1, 1, 0, 1, 0 ) \quad .
\end{equation}
Analogously, we construct a second filtered time series $\xi^{(\beta)}$ based on a second $\beta$-quantile. We then calculate the lagged cross-correlation 	of the filtered time series for each lag $l\in(-T/2,T/2)$
\begin{equation}
	{\rm qcf}_l(\xi^{(\alpha)}, \xi^{(\beta)}) = \frac{ 1 }{ T } \sum_{t=1}^{T - l} \frac{ (\xi_{t}^{(\alpha)} - \bar{\xi}^{(\alpha)}) (\xi_{t+l}^{(\beta)} - \bar{\xi}^{(\beta)}) }{ \sigma^{(\alpha)} \sigma^{(\beta)} }
	\label{eq:qcf2}
\end{equation}
where $\bar{\xi}^{(\alpha)}$ denotes the mean value of the filtered time series $\xi^{(\alpha)}$. The standard deviation of the filtered time series is denoted by $\sigma^{(\alpha)}$. The basic idea of quantile-based filtered binary time series was first put forward by~\cite{Kedem1980}. 

The case of equal probability levels in the time domain is discussed in~\cite{Linton2007} and
\cite{Hagemann2011} who proposed a discrete Fourier transform of the correlations in (\ref{eq:qcf2}) with fixed $\alpha= \beta$ in order to develop robust spectral analysis. For an alternative estimate and the case $\alpha \not = \beta$ see~\cite{Dette2011a}. \\
We calculate 95\% confidence intervals for the quantile correlation functions by taking the standard deviation of the fluctuations of the $(0.5,0.5)$ quantiles around zero and multiply them with the standard score of 1.96 corresponding to a $0.95$ confidence level. This mitigates problems which would be introduced using the standard approach for confidence intervals, \textit{i.e.}, $1.96$ divided by the square root of the sample size, which assumes that the elements of the time series are \textit{i.i.d}. This assumption is not valid for the financial data we consider in the following.

From the setup of the quantile correlation function we immediately see that for equal probabilities $\alpha=\beta$ Eq.~(\ref{eq:qcf2}) yields the autocorrelation of the filtered time series. In this case the quantile correlation function is symmetric, \textit{i.e.}, ${\rm qcf}_l(\xi^{(\alpha)}, \xi^{(\alpha)})={\rm qcf}_{-l}(\xi^{(\alpha)}, \xi^{(\alpha)})$. If the probabilities differ, $\alpha\neq\beta$, the quantile correlation function is not necessarily symmetric.

We now clarify the meaning of possible combinations for $\alpha$ and $\beta$. We denote the combination of probabilities with $(\alpha, \beta)$. For example, if we choose the $(0.05,0.05)$ quantiles, the filtered time series will each contain 5\% ones, which correspond to the smallest 5\% of values in the time series $x$ according to Eq.~(\ref{eq:qcf}). In this case, we correlate the smallest values of the time series in Eq.~(\ref{eq:qcf2}). On the other hand, consider the $(0.95,0.95)$ quantiles. Here, we correlate the 95\% of the smallest values of the time series $x$. However, in the case of financial time series it is more interesting to know how the largest 5\% of values are correlated. The $(0.95,0.95)$ quantiles also indicate this. We notice that if we change the less-than or equal to sign in Eq.~(\ref{eq:qcf}) to a greater-than sign for both filtered time series $\xi^{(\alpha)}$ and $\xi^{(\beta)}$ we get the complement of the filtered time series, \textit{i.e.}, ones become zeroes and zeroes become ones. Readers with a background in computer science will notice that this is equivalent to a binary NOT operation on each filtered time series. As long as this operation is performed on both filtered time series the sign of the quantile correlation function will not change, compare Eq.~(\ref{eq:qcf2}). This leads us to the interesting case where we want to know how the smallest 5\% values are correlated to the largest 5\%. We choose the $(0.05,0.95)$ quantiles and effectively correlate the smallest 5\% of the values with the smallest 95\%. To answer the question, we have to change the lesser-than or equal to sign only for the second filtered time series $\xi^{(0.95)}$. This results in a change of sign for the quantile correlation function. Suppose we find a negative correlation for the $(0.05,0.95)$ quantiles. This means that the smallest 5\% and 95\% of the time series are negatively correlated. At the same time it implies that the smallest 5\% and the largest 5\% of the time series are positively correlated. To keep the notation simple, we will always calculate the filtered time series according to Eq.~(\ref{eq:qcf}) and mention how to interpret the quantile correlation function in the given context.

\section{Empirical study}
We conduct an empirical study using intraday data from the New York Stock Exchange (NYSE) from 2007 and 2008. We restrain ourselves to stocks from the S\&P 500 index, which consists mostly of the largest corporations in the USA. These stocks are traded frequently, giving us enough trades per day for meaningful statistics. The time series for the trades are accurate to the second. We discard the first and last ten minutes of each trading day to minimize effects due to the closing and opening auction. This leads to a time series of $22200$ seconds per day, ensuring that the studied time series are solely the result of the continuous trading, which uses the double auction order book mechanism. If no trade takes place during a given second we use the previous price for this second.
The NYSE data set contains a huge amount of data which needs some preparation before using it. We only take into account regular orders and require that at least in $800$ different seconds trades have taken place on a day for a given stock. Otherwise we do not use the trades of this day. The quantile-based correlation is always calculated on these intraday time series of $22200$ seconds. This is necessary due to the interday gap between the closing price and the opening price of the next trading day. We average the quantile-based correlation functions over all available trading days of roughly $250$ days for each year. From the price series $S_k(t)$ of each stock $k$, we calculate the return time series 
\begin{equation}
	r_k(t)  = \frac{ S_k(t+\Delta t) - S_k(t) }{ S_k(t) }
\end{equation}
for one-minute returns $\Delta t = 60$s. For the following study we only take the return time series into account. We notice that the quantile correlation function is invariant under monotone transforms. Therefore the binary time series are unaffected by the choice of returns, \textit{i.e.}, non-logarithmic or logarithmic.

\subsection{Empirical data}
\begin{figure*}[htbp]
  \begin{center}
    \includegraphics[width=0.95\textwidth]{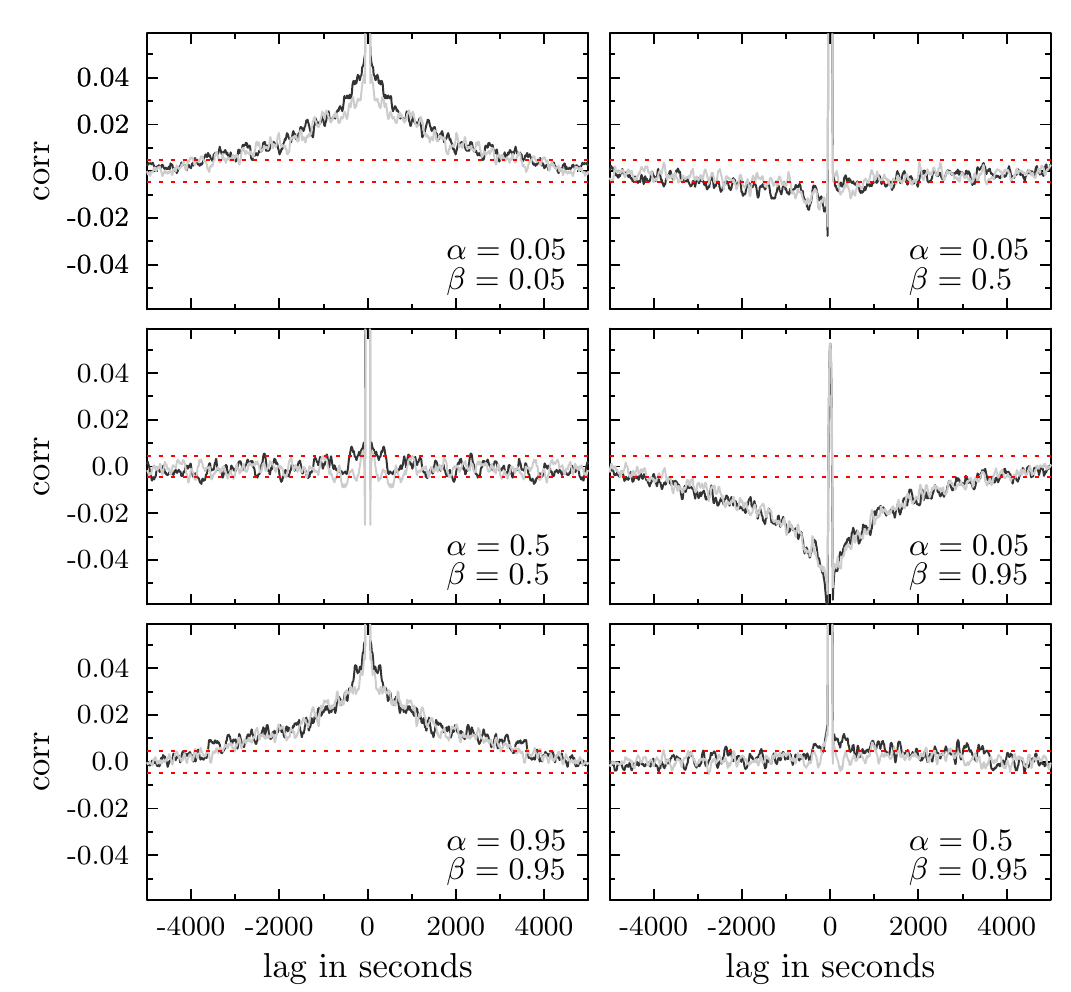}
  \end{center}
 \caption{Quantile correlation function for Abercrombie \& Fitch Co. (ANF) for 2007 (black) and 2008 (grey).}
 \label{fig:single}
\end{figure*}

We study the quantile-based correlation for six different quantile pairs $(\alpha, \beta)$.
Figure~\ref{fig:single} shows the quantile-based correlation for Abercrombie \& Fitch Co. (ANF). The $(0.05,0.05)$ quantiles correlate only the largest negative returns, while the $(0.95,0.95)$ quantiles indirectly correlate the largest positive returns. This requires further explanation. In principle, the $(0.95,0.95)$ quantiles correlate, according to Eq.~(\ref{eq:qcf}),  all returns which are smaller than the $0.95$-quantile. This is equivalent to the statement that the largest $5$\% of returns are correlated, because if we switch both lesser signs in Eq.~(\ref{eq:qcf}) to greater signs the quantile correlation function will not change as discussed in section~\ref{ch:qcf}. In both cases the correlation is non-zero and decays up to lags of roughly one hour to zero.  

\begin{figure*}[htbp]
  \begin{center}
    \includegraphics[width=0.95\textwidth]{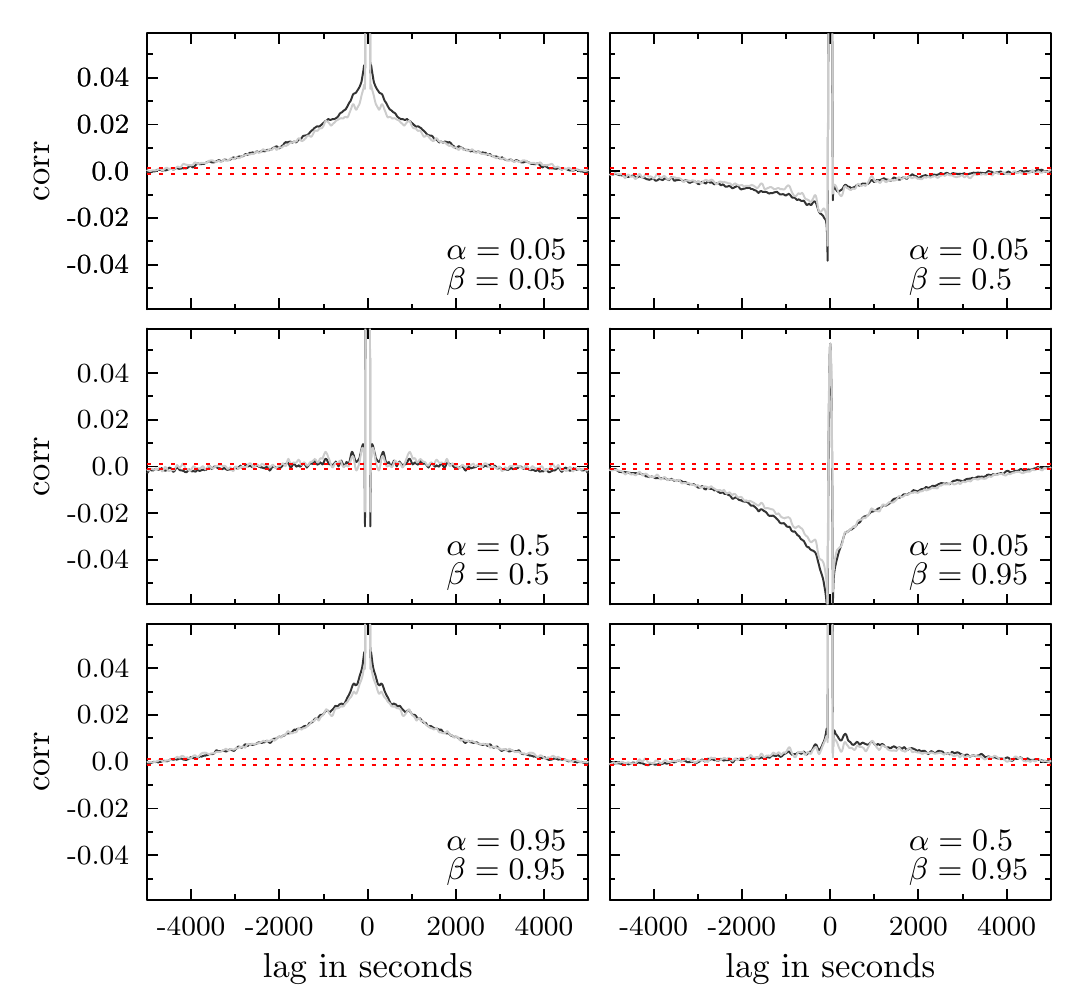}
  \end{center}
 \caption{Average quantile correlation function for $479$ stocks from the S\&P 500 index for 2007 (black) and $488$ stocks for 2008 (grey).}
 \label{fig:all}
\end{figure*}
We notice that the absolute values of the correlation are smaller to what is usually observed by using the autocorrelation of absolute or squared returns. This is due to the filtered time series which contain only zeroes and ones. Hence, the absolute correlation of these reduced time series is smaller. The $(0.5,0.5)$ quantiles correspond to the correlation of the sign of the returns if the distribution of returns has zero mean and is symmetric. As for the autocorrelation of returns this function should be zero, which is the case, \textit{i.e.}, all values are within the confidence interval and therefore not significant. Otherwise there would be arbitrage opportunities. For empirical return distributions, we cannot assume that the distribution of returns is perfectly symmetric. Hence, the shape of the correlation function for the $(0.5,0.5)$ quantiles differs within the confidence interval. For stochastic processes with symmetric return distributions we find zero correlation in section~\ref{ch:garch}

If the probabilities for the quantiles are chosen equally, \textit{i.e.}, $\alpha = \beta$, the quantile-based correlation functions must be symmetric for positive and negative lags. 

In contrast, for different quantiles, \textit{i.e.}, $\alpha \neq \beta$, we observe significant asymmetries. At first glance the asymmetry in the $(0.05,0.95)$ quantiles may be hard to spot, but a close look reveals that the area under the curve is smaller for positive lags. We calculate the normalized difference 
\begin{equation}
	\Delta A = \frac{ A_{-} - A_{+} }{ A_{-} + A_{+} }
\end{equation}
between the areas under the curve for negative and positive lags, where $A_{\pm}$ is
\begin{equation}
	A_{\pm} = \sum_{l=\pm 1}^{\pm T/2 } | {\rm qcf}_l(0.05, 0.95) | \quad .
\end{equation}
The measure lies in the interval $[-1,+1]$. For example, if the area under the curve is zero for negative lags and greater than zero for positive lags the measure is minus one. If the area under the curve is equally distributed between positive and negative lags, the normalized difference is zero. The results are shown in Table~\ref{tab:diff}. We find that the area under the curve for negative lags of the quantile correlation function is 8\% (2007) and 5\% (2008) larger compared to the area under the curve for positive lags. 
\begin{table*}[htbp]
\centering
\begin{tabular}{rrrrrrrr}
\hline
Figure & Dataset & Year & $\Delta A$ \\ 
\hline
\ref{fig:single} & ANF   & 2007 & 8\% \\
\ref{fig:single} & ANF   & 2008 & 5\% \\
\ref{fig:all} & SP500 & 2008 & 11\% \\
\ref{fig:all} & SP500 & 2008 & 5\% \\
\ref{fig:index} & INDEX & 2007 & 19\% \\ 
\ref{fig:index} & INDEX & 2008 & 6\% \\ 
\ref{fig:gjrgarch1} & GJR-GARCH & 2007 & 7\% \\ 
\ref{fig:gjrgarch1} & GJR-GARCH & 2008 & 1\% \\
\ref{fig:gjrgarch2} & GJR-GARCH & 2007 & 4\% \\ 
\ref{fig:gjrgarch2} & GJR-GARCH & 2008 & 1\% \\ 
\hline
\end{tabular}
\caption{Normalized difference $\Delta A$ of the area under the curve in the case of $(0.05,0.95)$ quantiles.}
\label{tab:diff}
\end{table*}

Here, we have to be careful with the interpretation of the probabilities $\alpha$ and $\beta$. What we see in Figure~\ref{fig:single} is the correlation of the smallest $5$\% of returns with the  smallest $95$\% of returns. This correlation is negative. If we want to correlate the smallest $5$\% with the largest $5$\% of returns we have to flip the sign of the quantile-based correlation function, because this is equal to change the second lesser sign in Eq.~(\ref{eq:qcf}) to a greater sign. Doing so will invert only the second filtered time series, which leads to a change of the sign of the quantile-based correlation function.
Therefore, we observe an asymmetry in the positive correlation of the smallest and largest returns. The slowly decaying correlation is reminiscent of the volatility clustering observed for equal probabilities $\alpha = \beta$. However, the asymmetry indicates the appearance of the leverage effect. 

Figure~\ref{fig:all} shows the average quantile-based correlation function for all stocks from the S\&P-500 index in the year 2007 (black) and 2008 (grey). The basic features remain the same compared to a single stock.

\begin{figure*}[!htb]
  \begin{center}
  
   \subfloat[Lag of 120 seconds] {
    \includegraphics[width=0.45\textwidth]{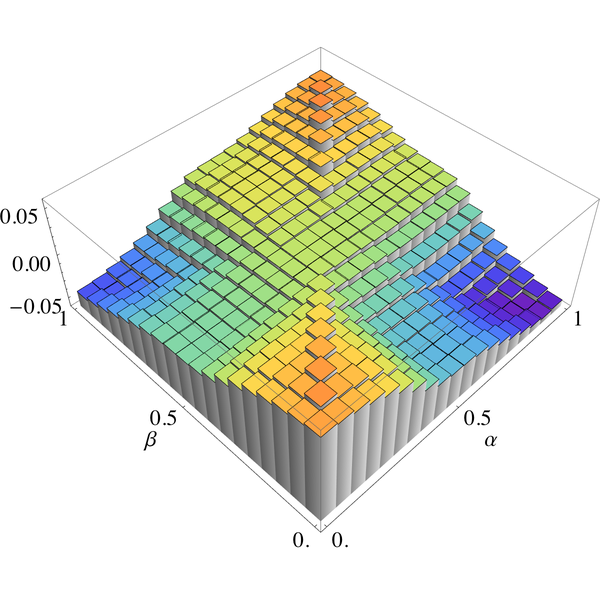}
\label{fig:qcf:a}
}
  \subfloat[Lag of 600 seconds] {
    \includegraphics[width=0.45\textwidth]{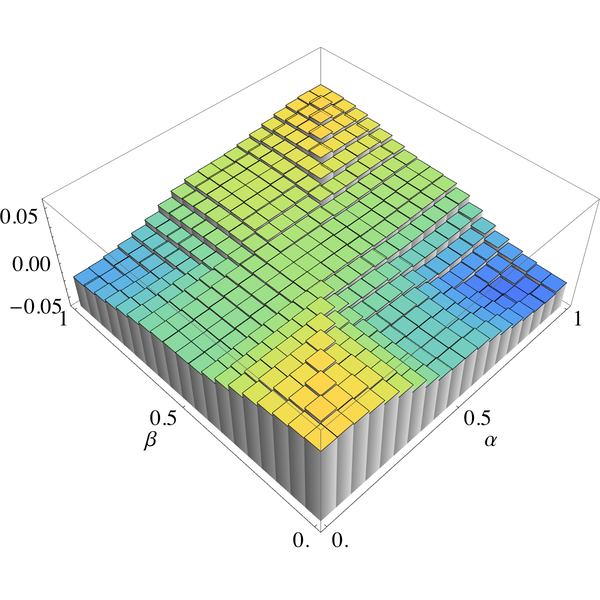}
\label{fig:qcf:b}
}

  \subfloat[Lag of 1200 seconds] {
    \includegraphics[width=0.45\textwidth]{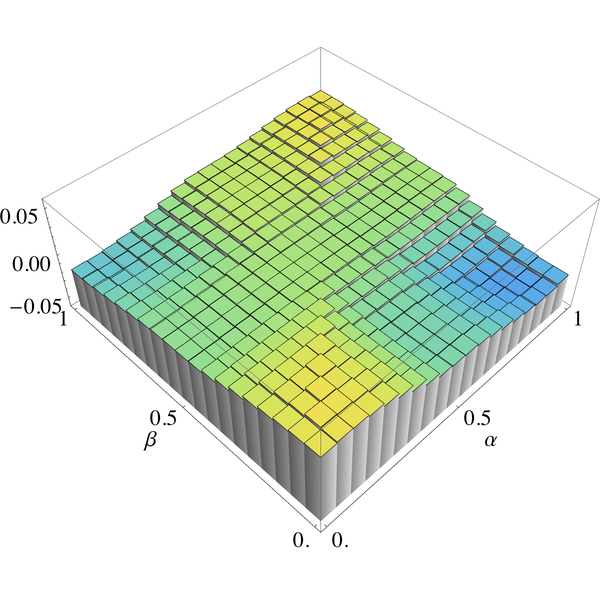}
\label{fig:qcf:c}
}
  \subfloat[Lag of 3600 seconds] {
    \includegraphics[width=0.45\textwidth]{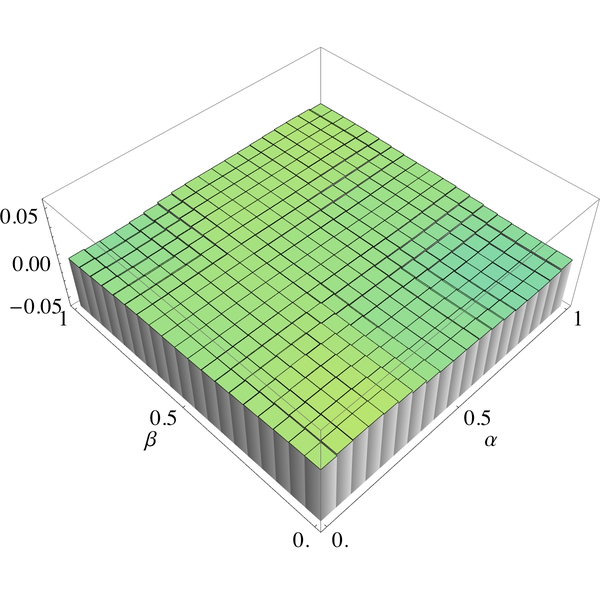}
\label{fig:qcf:d}
}
  \end{center}
 \caption{Average quantile correlation function for fixed lags calculated from $479$ stocks in the S\&P 500 during 2007 for four different lags.}
 \label{fig:qcf}
\end{figure*}

Another way to visualize the quantile correlation function is to look at a probability-probability plot for a fixed lag. The advantage is that it gives a more complete picture for different probability pairs with the caveat of only showing one lag. The result for all stocks from the S\&P 500 index in 2007, which corresponds to figure~\ref{fig:all}, are shown in figure~\ref{fig:qcf} for lags of 120, 600, 1200 and 3600 seconds. Importantly the plots also contain the information for the corresponding negative lags, because if we swap the probabilities $(\alpha,\beta) \rightarrow (\beta, \alpha)$ in equation~(\ref{eq:qcf2}) the lag axis also changes its sign $l \rightarrow -l$. The peaks for small probabilities around $(0.05,0.05)$ and large probabilities around $(0.95,0.95)$ are clearly visible and decay for larger lags. We also observe the asymmetries for probabilities $\alpha \neq \beta$ on the left and right hand side of the plot. Around the positive and negative peaks at the edges of the plot we observe plateau like areas.

\begin{figure*}[htbp]
  \begin{center}
    \includegraphics[width=0.95\textwidth]{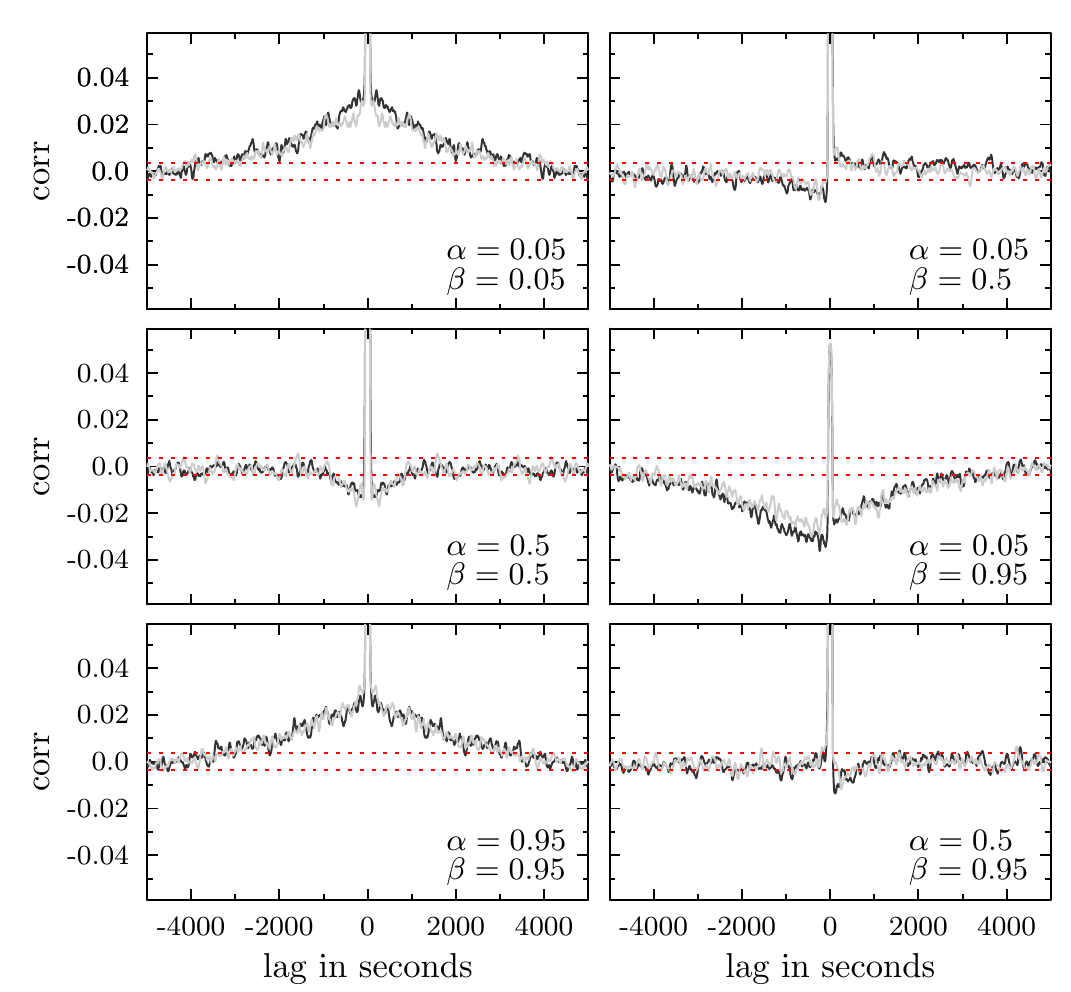}
  \end{center}
 \caption{Quantile correlation function for a equally weighted index calculated from the S\&P 500 stocks for 2007 (black) and 2008 (grey). }
 \label{fig:index}
\end{figure*}

For Figure~\ref{fig:index} we calculated a homogeneously weighted index from all stocks. We observe that the asymmetry for the $(0.05,0.95)$ quantiles becomes more pronounced. This behavior is connected to the ``correlation leverage effect'' studied by~\cite{Reigneron2011}. The authors find that the volatility of the index is comprised of the volatility of the single stocks and the average correlation between these stocks, which leads to stronger leverage effect for indices.  However, the absolute values of the anti-correlation are smaller compared to Figure~\ref{fig:all}.

\subsection{GARCH processes}
\label{ch:garch}

\begin{figure*}[htbp]
  \begin{center}
    \includegraphics[width=0.95\textwidth]{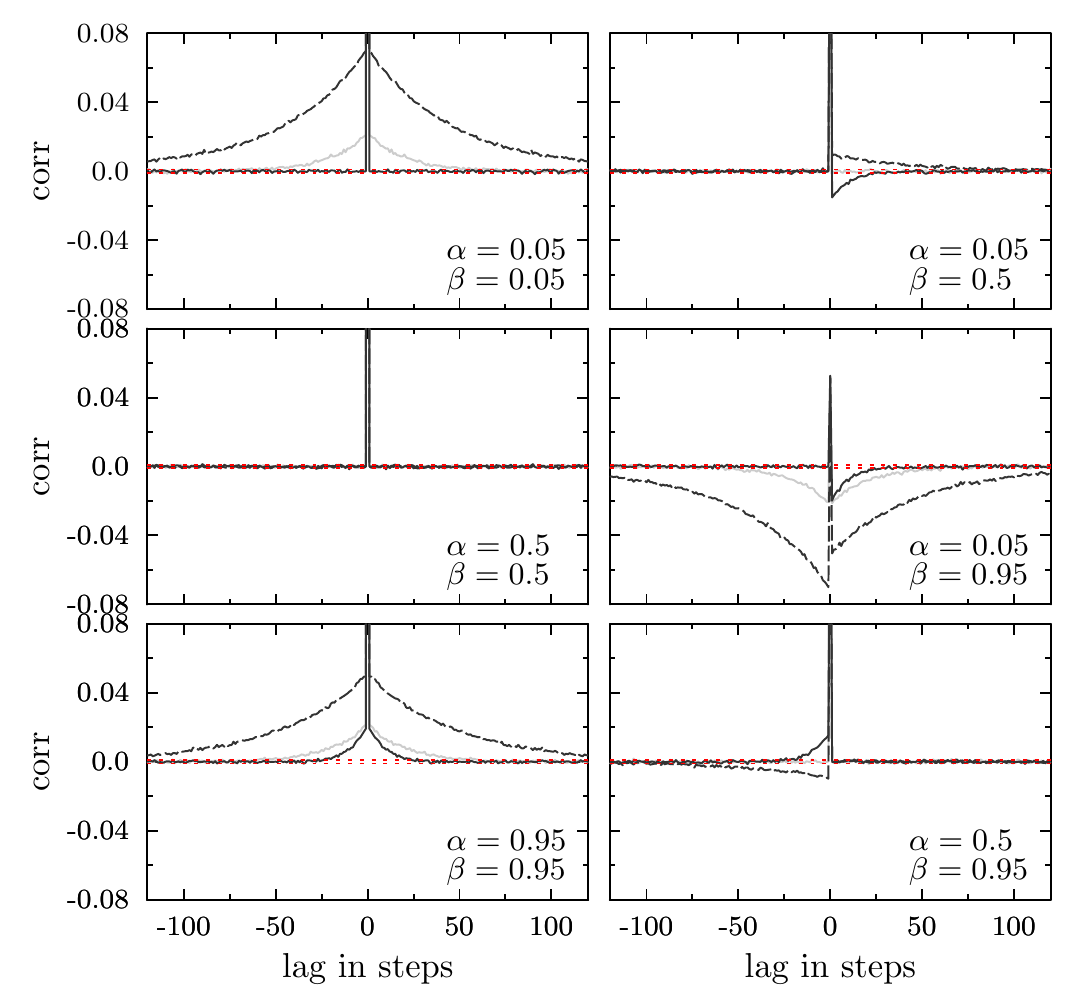}
  \end{center}
 \caption{Quantile correlation function for three stochastic processes GARCH (grey), GJR-GARCH (dashed) and EGARCH (black).}
 \label{fig:garch}
\end{figure*}
Figure~\ref{fig:garch} shows the quantile-based correlation function for three exemplary processes of the GARCH family. For all three cases we use GARCH processes of the order (1,1), see~\cite{Bollerslev2008} for a review.
The GARCH returns are modeled by
\begin{equation}
	\varepsilon_t = \sigma_t z_t \quad ,
\end{equation}
where $z_t$ is the stochastic part, \textit{i.e.}, a random variable drawn from a strong white noise process and the  conditional variances $\sigma_t^2$ are 
\begin{equation}
	\sigma_t^2 = \omega + \alpha_1 \varepsilon_{t-1}^2 + \beta_1 \sigma_{t-1}^2 \quad ,
\end{equation}
where $\omega, \alpha_1$ and $\beta_1$ are coefficients. The EGARCH(1,1) models the logarithmic variances according to
\begin{equation}
	\log \sigma_t^2 = \omega + \alpha_1( | z_{t-1}| - \langle | z_{t-1} | \rangle ) + \gamma_1 z_{t-1} + \beta_1 \log \sigma_{t-1}^2
\end{equation}
with the asymmetry parameter $\gamma_1$. Finally, the GJR-GARCH(1,1) uses
\begin{equation}
	\sigma_t^2 = \omega + \alpha_1 \varepsilon_{t-1}^2 + \gamma_1 \varepsilon_{t-1}^2 I(\varepsilon_{t-1} < 0 ) + \beta_1 \sigma_{t-1}^2
\end{equation}
with the indicator function $I(\cdot)$ for the conditional variances.

We choose the same parameters for all processes as far as possible with $\omega=0.00001$, $\alpha_1=0.05$, $\beta_1=0.9$ and drift $\mu = 0.001$. For the EGARCH and GJR-GARCH we choose an asymmetry parameter of $\gamma_1=-0.06$ and $\gamma_1=0.06$, respectively. The different sign for the asymmetry parameter is due to the construction of the EGARCH and GJR-GARCH. We emphasize that the parameters are chosen to receive pronounced characteristics for the quantile correlation. Fitting the process to empirical data will be investigated in the following sections. In accordance with the literature and the \texttt{rugarch}~package~\citep{Ghalanos2014} for R, we denote the GARCH parameters with $\alpha_1$ and $\beta_1$ and do not confuse them with the probabilities for the $(\alpha, \beta)$ quantiles. The classic GARCH process (grey) is symmetric by design. Unsurprisingly, we observe no significant asymmetries. We picked two common modifications to the classic GARCH, the EGARCH (black) and GJR-GARCH (dashed), which have an additional asymmetry parameter. The EGARCH process only shows a clustering of large positive returns and no correlation for small negative returns. For the $(0.05,0.95)$ quantiles only positive lags show a negative correlation, while negative lags have zero correlation. The GJR-GARCH shows clustering of large negative and positive returns and asymmetric non-zero correlations for the $(0.05,0.95)$ quantiles. This asymmetry is also observable in the absolute height of the quantile-based correlation function for the $(0.05,0.05)$ and $(0.95,0.95)$ quantiles. 

Here, the quantile-based correlation of the $(0.5,0.5)$ quantiles is really the correlation of the return sign, because the innovations for the GARCH processes are drawn from a normal distribution. Therefore, the distribution of returns is symmetric and the quantile-based correlation function is zero.

\begin{figure*}[!htb]
  \begin{center}
  
   \subfloat[GJR-GARCH, $l=2$ steps] {
    \includegraphics[width=0.45\textwidth]{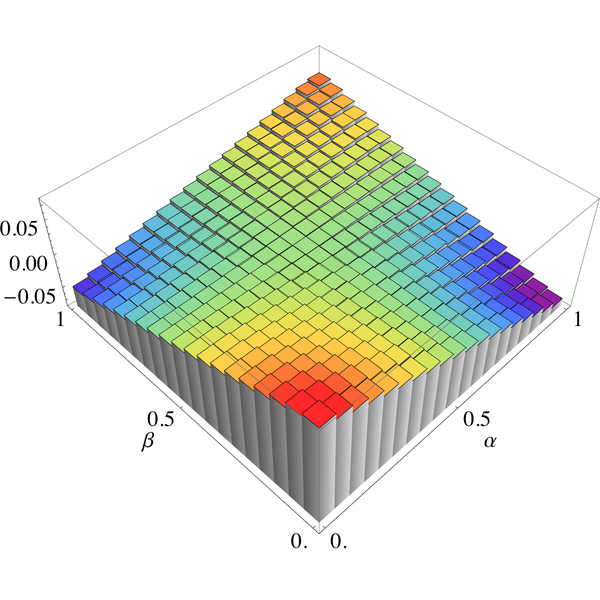}
\label{fig:qcf2:a}
}
  \subfloat[GJR-GARCH, $l=10$ steps] {
    \includegraphics[width=0.45\textwidth]{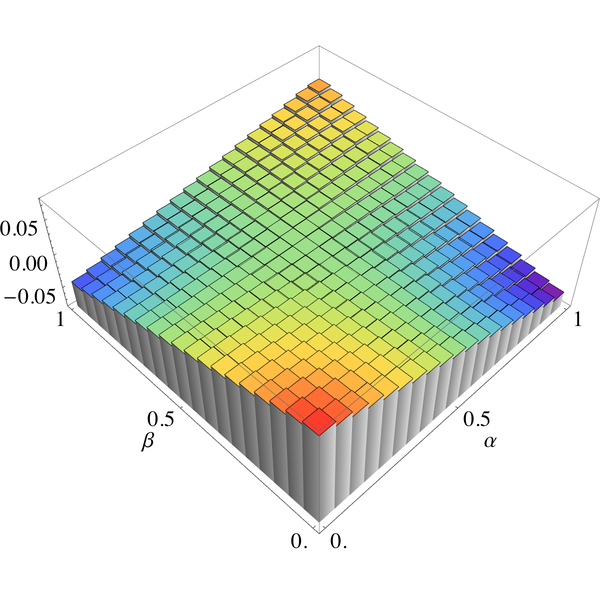}
\label{fig:qcf2:b}
}

  \subfloat[EGARCH, $l=2$ steps] {
    \includegraphics[width=0.45\textwidth]{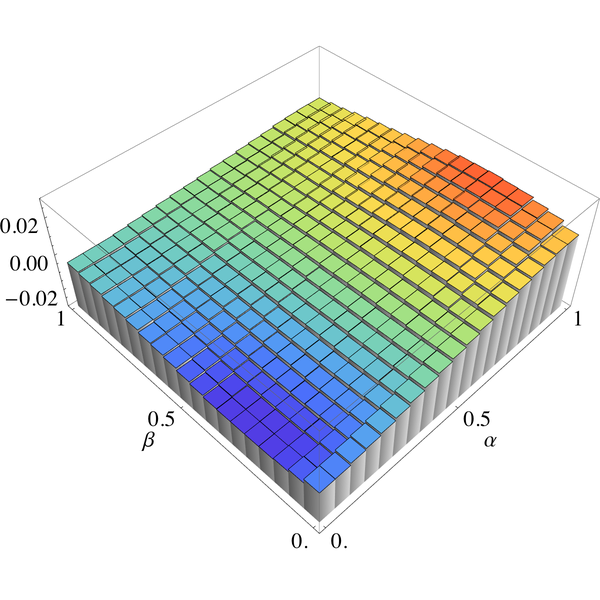}
\label{fig:qcf2:c}
}
  \subfloat[EGARCH, $l=10$ steps] {
    \includegraphics[width=0.45\textwidth]{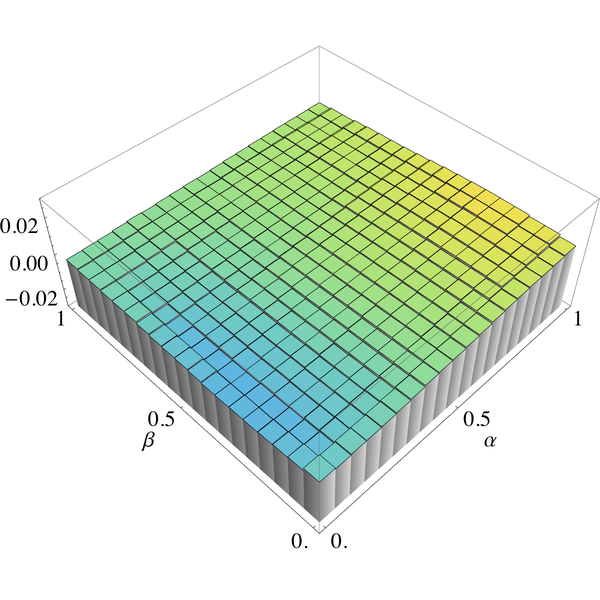}
\label{fig:qcf2:d}
}
  \end{center}
 \caption{Quantile correlation function for fixed lags calculated for the GJR-GARCH and EGARCH processes shown in figure~\ref{fig:garch}. Please note that we use different scales for the GJR-GARCH and EGARCH to better show their features.}
 \label{fig:qcf2}
\end{figure*}

In figure~\ref{fig:qcf2}, we present the probability-probability plot for the GJR-GARCH and EGARCH for two fixed lags. While the GJR-GARCH qualitatively captures the overall shape quite good, we find striking differences for the EGARCH. However, the GJR-GARCH does not reveal the plateau like structure we have seen for the average of the  S\&P 500 stocks. The plot for GJR-GARCH shows the peaks around the $(0.05,0.05)$ and $(0.95,0.95)$ probabilities and also the asymmetries around the $(0.05,0.95)$ probabilities. In contrast the EGARCH has a positive peak around the $(0.95,0.05)$ probabilities and a negative peak round the $(0.05,0.05)$ probabilities. 

While the asymmetric GARCH processes indeed show an asymmetric behavior in the correlation of very small and large returns the quantile correlation function uncovers differences to empirical data. For the $(0.05,0.5)$ and $(0.5,0.95)$ the GJR-GARCH and EGARCH show only non-zero behavior for the positive and negative lags, respectively. In addition, the EGARCH only shows non-zero correlations for positive lags for the $(0.05,0.95)$ quantiles. 

\subsubsection{Fitting each individual day}
\begin{figure*}[htbp]
  \begin{center}
    \includegraphics[width=0.95\textwidth]{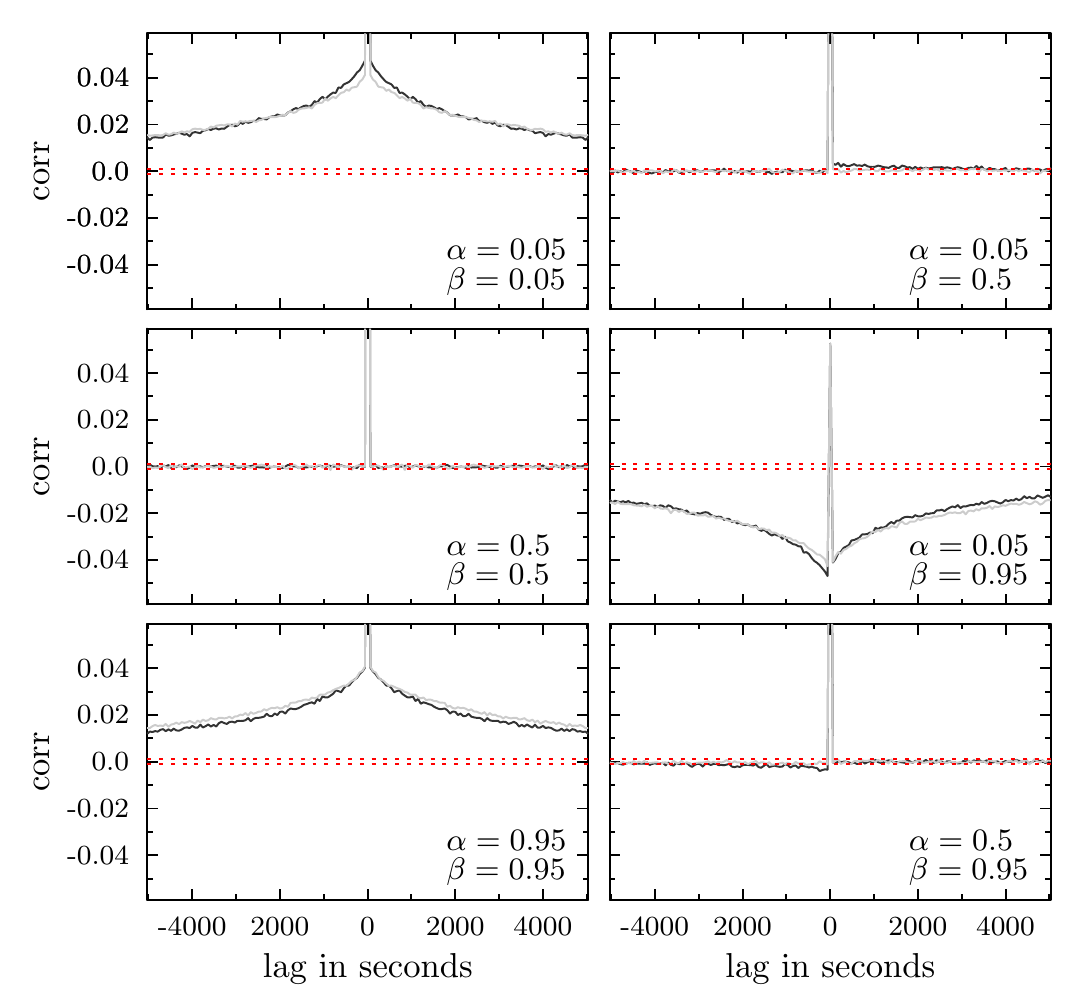}
  \end{center}
 \caption{Quantile correlation function for GJR-GARCH fitted to each trading day of the equally weighted S\&P 500 index for 2007 (black) and 2008 (grey).}
 \label{fig:gjrgarch1}
\end{figure*}
From the discussion in the previous section we believe that the GJR-GARCH is the best candidate of the three processes to describe the empirical data. Therefore, we fit the GJR-GARCH model to the equally weighted index of the S\&P 500 stocks for 2007 and 2008 for each trading day. This yields $250$ individual parameter sets $(\mu, \omega, \alpha_1, \beta_1, \gamma_1)$ per year. For each parameter set, we simulate a time series and calculate the quantile correlation function. The average for the years 2007 and 2008 is shown in figure~\ref{fig:gjrgarch1}. However, this approach does not yield results which agree with the empirical results for the index, see figure~\ref{fig:index}. We observe a much slower decay of the quantile correlation function for the $(0.05,0.05)$, $(0.95,0.95)$ and $(0.05,0.95)$ probabilities. The asymmetries are smaller for 2007, where the normalized difference for the area under the curve differs by $7$\% in contrast to $19$\% for the index and for 2008 it is negligible (1\%). For the $(0.05,0.5)$ and $(0.5,0.95)$ probabilities we find a speed of the decay which is similar to the empirical data at least for 2007. However, the qualitative shape of the quantile correlation function does not agree with the empirical data. The fit is performed on the non-overlapping one-minute returns, which shortens the time series from 22200 to 370 entries. We conjecture that the intraday time series are too short, which leads to poor fits and unrealistic parameter sets on some days which are biasing the results.

\subsubsection{Average parameters}
The first approach where we use $250$ time series generated from individual parameter sets obtained by fitting each day does not provide a satisfying accordance with the empirical data. Here, we generate one parameter set for each year by averaging the $250$ parameter sets from the previous section and than simulate $250$ time series for both sets each.
\begin{figure*}[htbp]
  \begin{center}
    \includegraphics[width=0.95\textwidth]{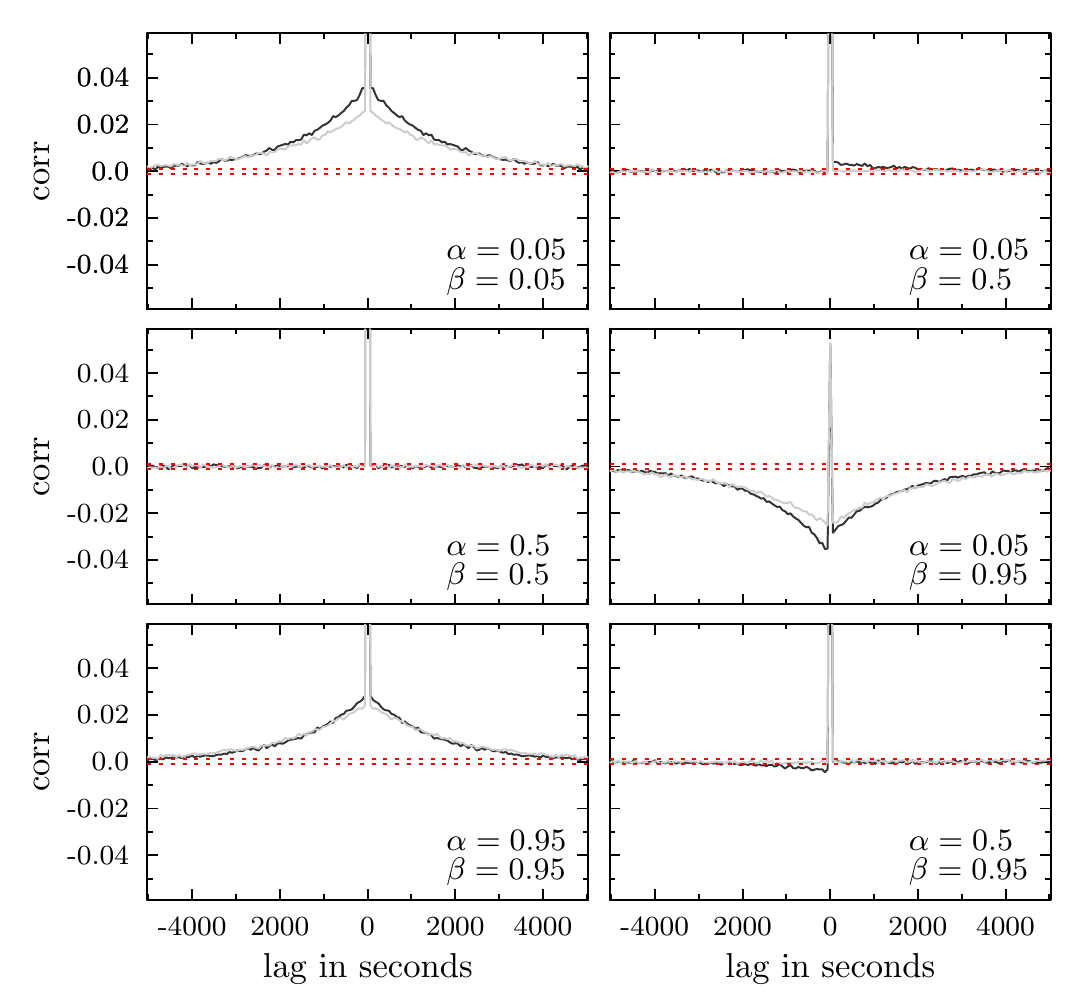}
  \end{center}
 \caption{Quantile correlation function for the GJR-GARCH process simulated from an averaged set of parameters received from the equally weighted S\&P 500 index for 2007 (black) and 2008 (grey).}
 \label{fig:gjrgarch2}
\end{figure*}
We calculate the quantile correlation function for each time series and average the results. For 2007 we find $\mu = -0.0008$, $\omega=0.0009$, $\alpha_1=0.0527$, $\beta_1=0.8986$ and $\gamma_1=0.0218$. In 2008 we find $\mu = -0.0026$, $\omega=0.0667$, $\alpha_1=0.0484$, $\beta_1=0.9191$ and $\gamma_1=0.0025$. At least for 2007 we are able to receive results which on an absolute scale reproduce the empirical data rather well. However, we still observe the qualitative discrepancies especially for the $(0.05,0.5)$ and $(0.5,0.95)$ probabilities. The asymmetry is clearly visible and we find a normalized difference between positive and negative lags of $4\%$. For 2008 we obtain a ten times smaller asymmetry parameter $\gamma_1$ in comparison to 2007. As the asymmetry parameter can be negative or positive between the trading days they can compensate to nearly zero. Thus, we observe no significant asymmetries for 2008.

Both approaches to fit the GJR-GARCH to the empirical intraday data fail to deliver satisfying results.

\section{Conclusion}
The quantile correlation function gives a detailed picture of the temporal dependencies in the underlying time series. It provides information which goes beyond the autocorrelation of the absolute or squared returns and uncovers asymmetries in the lagged correlations, which are connected to the leverage effect.

Beyond its usefulness for analyzing empirical time series it is a powerful tool to find subtle differences in time series obtained from stochastic processes which are designed to reproduce certain features. As an example we studied two stochastic processes of the GARCH family with asymmetry parameters and found differences compared to empirical data. Replicating all temporal features of return time series is crucial to achieve improved volatility forecasting. \cite{Martens2009} find that taking the leverage effect into account significantly improves the out-of-sample volatility forecast. \cite{Corsi2012} analyze among others a variation of the GJR-GARCH and further support the importance of including asymmetries. \cite{Hansen2005} notice that a GARCH(1,1) process is clearly inferior to models which account for the leverage effect with regard to the volatility forecast for stocks. In case of exchange rates they find no evidence that the GARCH(1,1) model is outperformed by more complex models. Numerous studies indicate that there is no definite answer to the question which stochastic process yields the best volatility forecast for financial data, see \cite{Poon2003} for an extensive review. In particular, there is no consensus whether EGARCH or GJR-GARCH performs best, see for example \cite{Bluhm2001}. The results largely depend on the data set under consideration, the time horizon, the test procedure and the stability of the fit. The quantile correlation function provides a means to further investigate why these deviating results occur for different data sets. It is beyond the scope of this paper to investigate every stochastic process, see~\cite{Bollerslev2008} for an overview of GARCH type processes. However, we advertize the reader to use the quantile correlation function to study his favorite stochastic process and its temporal dependencies in more detail. In general the quantile correlation function can serve as a sensitive tool to examine the agreement between stochastic processes and empirical time series.

\section{Acknowledgments}
The work of H. Dette has been supported in part by the
Collaborative Research Center ``Statistical modeling of nonlinear
dynamic processes'' (SFB 823, Teilprojekt A1 and C1) of the German Research Foundation
(DFG).

\clearpage


\end{document}